\documentclass[10pt, a4paper, twocolumn]{IEEEtran}
\usepackage[dvips]{color}
\usepackage{epsf}
\usepackage{times}
\usepackage{epsfig}
\usepackage{graphicx}

\usepackage{algorithm}
\usepackage{algpseudocode}

\usepackage{amsmath}
\usepackage{amssymb}
\usepackage{amsxtra}
\usepackage[numbers]{natbib}
\usepackage{enumerate}

\usepackage{physics}

\begin{document}

\title{\Large{COLLISION RESOLUTION AND INTERFERENCE ELIMINATION IN MULTIACCESS COMMUNICATION NETWORKS}}

\author{{\bf Naeem Akl, Ahmed Tewfik} \vspace{0.2cm} \\
Department of Electrical and Computer Engineering\\
UT Austin\\
\vspace{-0.78cm}}

\maketitle

\thispagestyle{empty}

\begin{abstract}
We define a multiaccess communication scheme that effectively eliminates interference and resolves collisions in many-to-one and many-to-many communication scenarios. Each transmitter is uniquely identified by a steering vector. All signals issued from a specific transmitter will be steered into the same single-dimensional or double-dimensional subspace at all receivers hearing this transmission. This subspace is orthogonal to the noise subspace at a receiver and the signals within the subspace can be extracted using the root-MUSIC method. At high SNR, local channel knowledge and strict synchronization, the algorithm asymptotically achieves full network capacity on condition that a channel remains constant within a single time slot. Without synchronization, the worst case asymptotic performance is still greater than the $50\%$ throughput achieved by collision resolution algorithms and interference management techniques like interference alignment. 

\end{abstract}
\textit{\textbf{Index --- collision resolution, interference elimination, interference alignment, root-MUSIC, steering vectors}} 

\section{Introduction}
  
Communication resources are scarce relative to the data requirements of the system users. A multiaccess communication scheme is needed to control the sharing of these resources among the different users. If a scheme allows several transmitters to communicate with the same receiver using the same resources, the scheme incurs collisions and retransmissions become necessary. In addition, if a scheme allows several transmitters to communicate with distinct receivers but a signal targeted to one receiver suffers from interference of signals targeted to others, this signal gets deteriorated and its decoding incurs errors if feasible in the first place. In both many-to-one and many-to-many communication scenarios, one category of schemes regulates multiaccess communication by preventing collisions and non-negligible interference. The other category advises collision-resolution protocols and signal-decoding algorithms in presence of interference.\\

\noindent Time and frequency division multiple access (TDMA and FDMA) are two conflict-free multiaccess schemes. A major drawback of both schemes is that only a portion of the available resources (time and frequency) is utilized when the network is lightly loaded while the other portion remains idle. Code division multiple access (CDMA) is another scheme that orthogonalizes the channel access via use of codes. Yet it is interference-limited in practice~\cite{multiaccess}. On the other hand, slotted Aloha, carrier sense multiple access(CSMA), and CSMA with collision detection and collision avoidance (CSMA/CD and CSMA/CA) are contention-based protocols. Transmitters with data compete to access the channel after random waiting periods while bearing the risk of collisions. This access mode is useful to accommodate variable bit rate data streams but is less efficient in heavily loaded networks. CSMA/CD is used in Ethernet and achieves a throughput of $1/e = 36.79\%$. It can be shown that an upper bound on the throughput of any collision resolution algorithm is $58.7\%$. Among other assumptions, this upper bound holds true for the case where all packets involved in a collision should be retransmitted~\cite{data_networks}. A throughput of $48.78\%$ is achieved using a tree algorithm for collision resolution~\cite{data_networks}.\\

\noindent Interference alignment, first introduced in~\cite{Khandani}, is a recent interference management technique that is neither resource-reservation-based nor contention-based. Instead, each transmitter steers its signal so that it lies within a reduced subspace along other interfering signals at every receiver except its desired receiver. Each receiver then looks outside its interference subspace and is able to extract the desired signal. For the fully connected $K$ user time-varying interference channel and assuming global channel knowledge, the authors in~\cite{Jafar} show that this channel has $K/2$ degrees of freedom which is a $50\%$ throughput. In~\cite{Jafar_distributed}, the authors study the feasibility of interference alignment given only local channel state information (CSI) at each node. Practical challenges for interference alignment are described in~\cite{Heath_practical}. Also, interference alignment is not designed for the many-to-one communication scenario since if all the transmitted signals need to be decoded by a receiver, no two signals may be aligned with respect to that receiver.\\

\noindent In this paper we describe an algorithm that works in both the many-to-one and many-to-many communication scenarios. Having said that, a receiver manages collisions and interference by decoding the collided packets and the interfering signals. Here we overlook security or privacy concerns whenever the decoder is not the target receiver. The channel is assumed packet-switched so that we avoid the inefficiencies of resource reservation under low load. In addition, while interference alignment and best collision resolution algorithms achieve a $50\%$ throughput, our algorithm asymptotically utilizes the full capacity of the network when the number of users grows infinite, provided that any communication (desired and undesired) between the transmitters and the receiver(s) is synchronized and the signal-to-noise ratio (SNR) is high. If the synchronization requirement cannot be met, the asymptotic performance is still $50\%$ in the worst case.\\  

\noindent We take a different perspective than interference alignment. In interference alignment, each receiver looks at its own space uniquely: its desired signal lies in half the space while all the other signals lie in the second half. The split is thus unique to every receiver. Since the transmitters have to steer their signals so that they get aligned as desired by the receivers, they become over-constrained and the steering achieves only $50\%$ efficiency. However, we let the receivers have a unified view of the signal space. This does not mean cooperation among the transmitters or the receivers. Instead, all the receivers view the space as two orthogonal subspaces: one subspace that holds all the signals and noise, and another subspace that holds only noise. The first subspace has dimensionality of the order of the number of transmitted signals. Under perfect synchronization this is exactly the number of transmitters. The second subspace has enough dimensions to suppress noise. This could be single-dimensional for high SNR. We choose the steering vectors to uniquely identify each transmitter. Since the number of dimensions occupied by the transmitted signals always grows as their number increases, the problem is no more overconstrained and every transmitted signal can be decoded by every receiver. If an arriving signal is so weak, a receiver can always decide that this is an interfering signal and move the occupied dimension(s) to the noise subspace.\\

\noindent We select the same steering vectors as those of the root-MUSIC algorithm used in~\cite{music} to estimate the signal direction of arrival in an array antenna system. For our algorithm, since these vectors lie in a high dimensional subspace, it is indifferent whether the extension over the dimensions occurs in time, frequency or space. In this paper we assume the channel is a single carrier and every node has a single antenna, so the steering vectors extend over time. This paper is focused on describing the algorithm. Network performance analysis and optimizations on the functionality of the algorithm will be excluded. The algorithm is defined incrementally over sections~\ref{sec3} to~\ref{sec8}, where each increment considers a more general or realistic setting. We first present the system model.

\section{System Model}~\label{sec2}
Consider a single-carrier system in which $K$ transmitters contact a single receiver, each node having a single antenna. Each transmitter accesses the channel whenever data is available without waiting for the channel to be idle. We assume there is only a single receiver because a receiver is able to decode all arriving signals without cooperation with other receivers. Thus in a system with more than one receiver, they all perform the same functionality. Obviously each receiver will be able to only decode its unique set of arriving signals, whether desired or not. The minimum transmission unit is a packet with $P$ symbols that could be real or complex. It takes one time slot to transmit a single packet. A symbol duration is $\tau$ seconds, so $1~\text{slot} = P\times \tau$. We are interested in the case where $K>1$. For $K=1$, we assume that the transmitter identifies immediately upon sending the packet that there were no concurrent transmissions and there is no need for further action. This could be through immediate feedback or simply the transmitter can sense the channel. This assumption is not absolutely necessary nor critical for the network asymptotic throughput but is made for simplicity. Note that a feedback from the receiver is only possible if the packet has extra bits for error detection. We refer to these bits as CRC for cyclic redundancy check.\\

\noindent On the other hand, if there are concurrent transmissions, each transmitter will send its packet more than once before the receiver can decode these individual packets. During this time, some new transmitters might join and start sending new packets. The receiver in turn will need more time to decode all the received packets. In such a scenario, $K$ refers to the total number of transmitters at the instant of successful decoding, which is greater than or equal to the number of transmitters upon arrival of the first few packets. The time for successful decoding is measured in slots and is equal to $N$. This will be function of $K$, the SNR and the complexity of the communication scenario (like when each transmitter joined the set of active transmitters). We will not consider the impact of the last factor on $N$ in this paper and assume that the receiver has high processing capabilities. We also assume that the receiver will try to minimize $N$ and in a sense also minimize $K$. For instance, if there is a single transmission in time slot 1 and another transmission from a different receiver in time slot 2, the receiver will detect these two transmissions as two separate scenarios, each with $N=1$ and $K=1$. Note that the receiver does not know $K$ beforehand. Since data availability at transmitters is random, then so are $N$ and $K$.\\

\noindent All vectors $\overrightarrow{v}$ are column vectors and have arrow symbols on top. The transpose of $\overrightarrow{v}$ is $\overrightarrow{v}^T$ and the conjugate transpose of $\overrightarrow{v}$ is $\overrightarrow{v}^H$. Similar transpose notation is used for matrices. If $\overrightarrow{v}$ has length $L$, then $\overrightarrow{v}[l]$ is the $l^{\text{th}}$ element of $\overrightarrow{v}$, $1\leq l\leq L$. Define
\begin{equation*}
\overrightarrow{v}^{(d)} = \begin{cases}
\big[\overbrace{0, \dots, 0}^{d~\text{zeros}}, \overrightarrow{v}[1], \dots, \overrightarrow{v}[L-d]\big]^T,~~1\leq d\leq L-1\\
\quad\quad\quad\quad\overrightarrow{v}\quad\quad\quad\quad\quad\quad\quad\quad\quad~,\quad d=0\\
\big[\overrightarrow{v}[1-d], \dots, \overrightarrow{v}[L],\underbrace{0, \dots, 0}_{-d~\text{zeros}}\big]^T,~~1-L\leq d\leq -1
\end{cases}
\end{equation*}

\noindent A packet is represented as a vector of symbols, were $\overrightarrow{s}_k$ is the packet to be transmitted by transmitter $k$, $1\leq k\leq K$. During time slot $n$, $1\leq n\leq N$, the channel between transmitter $k$ and the receiver remains constant and can be represented by a single complex number $h_k(n)$. This is one criterion to choose $P$. We assume the receiver knows $h_k(n)$ for every desired or undesired transmitter $k$ currently heard by the receiver during all slots $n$. This is only local channel knowledge in the sense that the receiver does not need to collect CSI for other receivers as in interference alignment. The noise over all channels is complex normal $\mathcal{CN}(0,\sigma^2 I)$ of mean 0 and covariance $\sigma^2 I$. A collection of $p\times q$ noise samples is denoted as $\mathcal{N}_{p,q}$. Each transmitter $k$ that might contact the receiver is assigned a unique complex exponential $r_k = e^{j\angle r_k}$ lying on the unit circle, where $0\leq \angle r_k < \pi$. The receiver is also aware of this assignment. Define the $N$-time extension of the steering vector of transmitter $k$ as
\begin{equation}
\overrightarrow{w}_{k,N} = \big[r_k^0, r_k^1,\dots, r_k^{N-1}\big]^T
\end{equation}
\noindent During time slot $t$ and whenever at least one transmitter is active, the receiver collects a vector of symbols
\begin{equation}
\overrightarrow{y}_n = \big[\overrightarrow{y}_n[1], \overrightarrow{y}_n[2],\dots, \overrightarrow{y}_n[P]\big]^T
\end{equation} 

\section{Alignment at time $t=0$}~\label{sec3}
Assume in this section that $h_k(n) = 1$ for all $k$ and $n$ and the SNR is high. Assume also that all transmitters are synchronized together and with the receiver so that the $K$ transmitted packets happen to arrive at the receiver at exactly time $t=0$, i.e. the start of slot $n=1$. The receiver checks the CRC of $\overrightarrow{y}_1$. For $K>1$, this will be corrupted since $\overrightarrow{y}_1 = \sum_{k=1}^{K}\overrightarrow{s}_k + \mathcal{N}_{P,1}$, which is erroneous at high SNR due to collision. The transmitters detect the collision, and each transmitter $k$ sends packet $r_k^1 \times \overrightarrow{s}_k$ that will exactly fit within slot $n=2$. Assume the receiver still fails to decode the packets $\overrightarrow{s}_1,\dots,\overrightarrow{s}_K$ after having collected by now vectors $\overrightarrow{y}_1$ and $\overrightarrow{y}_2$. During time slot $n$, each transmitter $k$ sends packet $r_k^{n-1}\times \overrightarrow{s}_k$. By the end of slot $n=N$ the receiver will have stacked $N$ vectors $\overrightarrow{y}_n^T$ horizontally into a single matrix $Y_N$ given by
\begin{equation}
\label{eqY_Ndetailed}
Y_N = \begin{pmatrix}
\overrightarrow{y}_1^T\\
\vdots\\
\overrightarrow{y}_N^T
\end{pmatrix} = \begin{pmatrix}
\overrightarrow{w}_{1,N} & \hdots & \overrightarrow{w}_{K,N}
\end{pmatrix} \times \begin{pmatrix}
\overrightarrow{s}_1^T\\
\vdots\\
\overrightarrow{s}_K^T
\end{pmatrix} + \mathcal{N}_{N,P}
\end{equation}
\noindent or shortly
\begin{equation}
\label{eqY_N}
Y_N = W_N\times S + \mathcal{N}_{N,P}
\end{equation}
\noindent Note that $W_N$ is a Vandermonde matrix. $P$ is chosen such that $P>K$. Consider the case when $N>K$. $W_N$ will have a non-trivial left nullspace of dimension $N-K$. Let $U_{\bot}$ hold as columns the basis vectors of the left nullspace of $W_N$. Multiplying equation~(\ref{eqY_N}) by $U_{\bot}^H$ from the left and observing that $\mathcal{N}_{N,P}\sim\mathcal{CN}(0,\sigma^2 I_{N\times N})$ we have 
\begin{equation}
\label{leftNull}
U_{\bot}^H Y_N \sim \mathcal{CN}(0,\sigma^2 I_{(N-K)\times(N-K)})
\end{equation}
\noindent by noting that $U_{\bot}^H U_{\bot} = I$. On the limit $\sigma^2\rightarrow 0$ (i.e. the SNR is very high),~(\ref{leftNull}) implies that $U_{\bot}^H$ converges in probability to the left nullspace of $Y_N$. Suppose this is the case. By applying the singular value decomposition (SVD) to $Y_N$ we have
\begin{equation}
Y_N = \begin{pmatrix}[U_{\parallel} & U_{\bot}]\end{pmatrix}\Sigma V^H
\end{equation}
\noindent $U_{\bot}$ then holds the column vectors in the left-singular matrix of $Y_N$ corresponding to the $N-K$ almost zero singular values. This suggests the following method for the receiver to decode the packets.\\

\noindent The receiver collects vectors $\overrightarrow{y}_n$ iteratively and continuously checks for the rank of matrix $Y_n$. Since the SNR on all channels $k$ is high, the receiver is able to zero-threshold the singular values of $Y_n$ that are solely due to noise. Thus, as long as $Y_n$ is full rank, the left nullspace of $Y_n$ is still trivial and an extra vector $\overrightarrow{y}_{n+1}$ needs to be collected. At $n=K+1$, the nullspace of $Y_n$ is no more trivial. This is because $W_N\times S$ in~(\ref{eqY_N}) has rank $K$ for all $N>K$ and thus the newly computed singular value at $n=K+1$ is only due to noise (thus very small at high SNR) and is thresholded to zero. This way the receiver now knows the number of transmitters $K$. Depending on the SNR, the receiver may choose to collect extra vectors $\overrightarrow{y}_{n}$ by not broadcasting any acknowledgment to the transmitters. These additional vectors provide extra dimensions to approximate the noise-only subspace of basis $U_{\bot}$. The receiver stops at $n=N>K$ and $U_{\bot}$ has $N-K$ column vectors.\\

\noindent Since the SNR is high, $U_{\bot}$ computed from the SVD of known matrix $Y_N$ is also roughly the left nullspace of unknown matrix $W_N$. As in~\cite{music}, define the steering vector $\overrightarrow{w'}_N$
\begin{equation}
\label{z_steer}
\overrightarrow{w'}_N = \big[1, z^{1}, \dots, z^{N-1}\big]^T
\end{equation}  
\noindent and solve for unknown complex number $z$ in equation
\begin{equation}
J(z) = \overrightarrow{w'}_N^H\times U_{\bot} U_{\bot}^H\times \overrightarrow{w'}_N = 0
\end{equation}
\noindent The closest $K$ solutions for $z$ to the unit circle and the unit complex exponentials $r_k$ and of angle $0\leq \angle z < \pi$ indicate to the receiver the identity of the $K$ transmitters. The receiver in turn constructs matrix $W_N$ using the $K$ identified exponentials $r_k$. The matrix of decoded packets is then given by
\begin{equation}
\hat{S} = (W_N^H W_N)^{-1} W_N^H Y_N
\end{equation}
\noindent Note that $(W_N^H W_N)^{-1}$ is full rank and thus admits an inverse. Note also that the order of columns of constructed matrix $W_N$ is unimportant since this will only affect the order of the decoded packets (rows) in matrix $\hat{S}$ without losing the identities of their transmitters.\\

\noindent If it happens that the signal power on some channel $k$ is small compared to the noise variance $\sigma^2$, the receiver might zero-threshold the corresponding singular value in the SVD of $Y_N$. In this case the signal on channel $k$ will be lying within the noise subspace $U_{\bot}$. This should not be harmful as long as the signal is undesired by the receiver. Note that $U_{\parallel}^H U_{\bot} = 0$ which establishes the orthogonality between the signal-and-noise subspace and the noise-only subspace of the received matrix $Y_N$. The fact that each signal (+ noise) occupies a separate dimension in $U_{\parallel}$ justifies the decodability of the $K$ packets. On condition that $N-K\sim\mathcal{O}(1)$ which is true at high SNR, the asymptotic throughput is
\begin{equation}
\label{throughput}
\lim_{K\rightarrow \infty}\frac{K}{N} = 100\%
\end{equation}

\section{Alignment at the Start of a Time Slot}~\label{sec4}
We still assume the same channel conditions as in Section~\ref{sec3}. We simplify the synchronization requirements and do not necessitate that all packets should arrive at the same instant $t=0$. However, in this section we assume the transmitters are synchronized with the receiver so that a packet from a transmitter arrives only at exactly the start of a time slot $n$, i.e. $t = (n-1)P\tau, 1\leq n\leq N$. An example scenario would be that packets from transmitters $1$, $2$ and $3$ arrive at the receiver at $t=0$. Excluding the noise terms for brevity, $\overrightarrow{y}_1 =  \overrightarrow{s}_1 + \overrightarrow{s}_2 + \overrightarrow{s}_3$. The transmitters detect the collision and retransmit their packets (now weighted) and the receiver collects $\overrightarrow{y}_2 =  r_1\overrightarrow{s}_1 + r_2\overrightarrow{s}_2 + r_3\overrightarrow{s}_3$. Matrix $Y_2$ still has a trivial nullspace and the receiver cannot decode the three packets. Now transmitter 4 has a packet to send. Transmitter 4 does not need to know it is 2 slots behind transmitters 1,2 and 3. Moreover, transmitters 1,2 and 3 do not recognize that transmitter 4 has joined the set of active transmitters. Still all transmissions will arrive at $t=(3-1)P\tau$. The receiver then collects $\overrightarrow{y}_3 =  r_1^2\overrightarrow{s}_1 + r_2^2\overrightarrow{s}_2 + r_3^2\overrightarrow{s}_3 + \overrightarrow{s}_4$. At $t=(4-1)P\tau$, $\overrightarrow{y}_4 =  r_1^3\overrightarrow{s}_1 + r_2^3\overrightarrow{s}_2 + r_3^3\overrightarrow{s}_3 + r_4\overrightarrow{s}_4$ and so on. In an arbitrary scenario, if the first packet from transmitter $k$ arrives at $t =(n_k-1)P\tau$, where $1\leq n_k\leq N-1$ for $1\leq k\leq K$, equation~(\ref{eqY_Ndetailed}) generalizes to
\begin{equation}
\label{eqY_N2}
\begin{split}
Y_N &= \begin{pmatrix}
\overrightarrow{w}_{1,N}^{(n_1-1)} & \hdots & \overrightarrow{w}_{K,N}^{(n_K-1)}
\end{pmatrix} \times S + \mathcal{N}_{N,P}\\
&= W_N^{(n-1)}\times S + \mathcal{N}_{N,P}
\end{split}
\end{equation}
\noindent where we abuse notation and choose $W_N^{(n-1)}$ for the steering matrix. The number of independent columns in $W_N$ and $W_N^{(n-1)}$ is the same and thus the receiver can still detect the number of transmitters $K$ in the same manner as in Section~\ref{sec3}. $Y_N$ and $W_N^{(n-1)}$ continue to have the same left nullspace at high SNR. $Y_N$ can still be split into two orthogonal subspaces: the signal-and-noise subspace and the noise-only subspace. The only change in the decoding algorithm illustrated in Section~\ref{sec3} is that the receiver needs to search for the K solutions closest to the unit circle and the unit complex exponentials $r_k$ among all the solutions of a whole set of independent equations
\begin{equation}
\label{eqSet}
J(z)^{n_k-1} = (\overrightarrow{w'}_N^{(n_k-1)})^H\times U_{\bot} U_{\bot}^H\times \overrightarrow{w'}_N^{(n_k-1)} = 0
\end{equation} 
\noindent for $1\leq n_k\leq N-1$. The upper bound on $n_k$ is $N-1$. This is because the receiver will not stop at a time slot in which a new packet arrives for the first time as it needs at least one final time slot to acquire a basis $U_{\bot}$ of a non-trivial left nullspace. Another way to see it is that $\overrightarrow{w'}_N^{(n_k-1)}$ for $n_k = N$ cannot be a steering vector since it has no root $z$ as in~(\ref{z_steer}). Based on what equation in set~(\ref{eqSet}) generates the $k^{\text{th}}$ solution, the receiver can also recover the shifts $n_k-1,~1\leq k\leq K$ and thus is able to build matrix $W_N^{(n-1)}$. The decoded set of symbols is then given by
\begin{equation}
\hat{S} = ((W_N^{(n-1)})^H W_N^{(n-1)})^{-1} (W_N^{(n-1)})^H Y_N
\end{equation}  
\noindent Note that the algorithm illustrated in Section~\ref{sec3} is a special case of the one in this section where $n_k=1$ for all $k$. It also achieves the same throughput as in~(\ref{throughput}). 

\section{Misaligned Transmissions}~\label{sec5}
Suppose not all transmitters can be synchronized with every receiver and thus it is not necessary that every packet arrives at the receiver at the start of a time slot. While synchronization is desired because of the full asymptotic throughput it achieves in~(\ref{throughput}), we now assume that a packet from transmitter $k$ may arrive during slot $n$ at $t=(nP+p')\tau$ where $0\leq p'\leq P-1$. For now $p'$ is an integer, i.e. the packet arrives only at the start of a symbol duration. We remove this restriction at the end of the section. Consider the following scenario. Packets from several transmitters arrive at $t=0$ so the receiver in the upcoming time slots is busy. Only the packet $\overrightarrow{s}_k$ of transmitter $k$ happens to arrive at $t = (0\times P+p)\tau$ for some integer $0<p<P$. Since this is the first transmission of packet $\overrightarrow{s}_k$ it is weighted by unity. Transmitter $k$ detects a collision and at $t = (1\times P+p)\tau$ packet $r_k\overrightarrow{s}_k$ arrives at the receiver. At $t=(2\times P+p)\tau$ packet $r_k^2\overrightarrow{s}_k$ arrives at the receiver and so on. Thus although the packets arriving from transmitter $k$ are misaligned, transmitter $k$ is unaware of that and behaves normally like all the other transmitters.\\

\noindent Since $\overrightarrow{y}_n$ defines the received $P$ samples within slot $n$, notice that the contribution of transmitter $k$ to $\overrightarrow{y}_1$ is $\overrightarrow{s}_k^{(p)}$. On the other hand, its contribution to $\overrightarrow{y}_2$ is $\overrightarrow{s}_k^{(p-P)} + r_k\overrightarrow{s}_k^{(p)}$. In slot $2$ the contribution to $\overrightarrow{y}_3$ is $r_k\overrightarrow{s}_k^{(p-P)} + r_k^2\overrightarrow{s}_k^{(p)}$. In slot $n>1$, transmitter $k$ contributes $r_k^{(n-2)}\overrightarrow{s}_k^{(p-P)} + r_k^{(n-1)}\overrightarrow{s}_k^{(p)}$ to $\overrightarrow{y}_n$. Therefore, from the perspective of the receiver, transmitter $k$ with original packet $\overrightarrow{s}_k$ and misaligned transmissions relative to the start of a time slot is equivalent to two transmitters $k_a$ and $k_b$ with original packets $\overrightarrow{s}_k^{(p)}$ and $\overrightarrow{s}_k^{(p-P)}$ and correctly aligned transmissions. Hypothetical transmitters $k_a$ and $k_b$ apply the same steering vector, and transmitter $k_a$ starts transmitting one time slot ahead of transmitter $k_b$. Therefore, in a set of $K$ active transmitters, each {\it misaligned transmitter} occupies two columns of $W_N^{(n-1)}$ and two rows of $S$ in (\ref{eqY_N2}), while an {\it aligned transmitter} simply occupies one column of $W_N^{(n-1)}$ and one row of $S$. Let $p_k+1$ be the index of the symbol at which $\overrightarrow{s}_k$ (or some weighted version) arrives within a time slot, $0\leq p_k<P$ and $1\leq k\leq K$. Define
\begin{equation}
\overrightarrow{s}_{k,p_k} = \begin{cases}
\overrightarrow{s}_k,\quad\quad\quad p_k = 0\\
\big[\overrightarrow{s}_k^{(p_k)},\overrightarrow{s}_k^{(p_k-P)}\big],~~0<p_k<P
\end{cases}
\end{equation}
\begin{equation}
\overrightarrow{w}_{k,N,p_k}^{(n_k-1)} = \begin{cases}
\overrightarrow{w}_{k,N}^{(n_k-1)},\quad\quad\quad p_k = 0\\
\big[\overrightarrow{w}_{k,N}^{(n_k-1)},\overrightarrow{w}_{k,N}^{(n_k)}\big],~~0<p_k<P
\end{cases}
\end{equation}
\noindent $Y_N$ can be expressed as
\begin{equation}
\label{eqY_Ndetailed2}
\begin{split}
Y_N &= \begin{pmatrix}
\overrightarrow{w}_{1,N,p_1}^{(n_1-1)} & \hdots & \overrightarrow{w}_{K,N,p_K}^{(n_K-1)}
\end{pmatrix} \times \begin{pmatrix}
\overrightarrow{s}_{1,p_1}^T\\
\vdots\\
\overrightarrow{s}_{K,p_K}^T
\end{pmatrix} + \mathcal{N}_{N,P}\\
&= W_{N,p}^{(n-1)}\times S_p + \mathcal{N}_{N,P}
\end{split}
\end{equation}
\noindent The columns of $W_{N,p}^{(n-1)}$ and the rows of $S_p$ continue to be linearly independent though their numbers are at least as those of $W_N^{(n-1)}$ and $S$ respectively. This implies $N$ is expected to increase compared to a scenario like that of Section~\ref{sec5}. The receiver iteratively checks at every slot whether $Y_N$ no more has a trivial left nullspace. This way the receiver can identify the total number of transmitters, both real and virtual (remember that each misaligned transmitter is replaced with two hypothetical ones). In addition, the receiver solves the set of equations~(\ref{eqSet}) and checks the number of unique roots close to the unit circle and set $\{r_k\}$. Since two hypothetical transmitters use the same steering vector, this latter number is the number of actual transmitters. Consequently, and using the difference of the last two computed counts, it is trivial for the receiver to find the number of each of the aligned and misaligned real transmitters.\\

\noindent Two hypothetical transmitters start transmitting in two consecutive slots using the same steering vector. It becomes tempting now to let the receiver also check which computed unit-magnitude solutions are repeated in two consecutive equations of set~(\ref{eqSet}) and which ones show only once. This way the receiver identifies which real transmitters are aligned and which are misaligned, builds $W_{N,p}^{(n-1)}$ and recovers $S_p$. However, there is a hidden obstacle to doing that. Let $\overrightarrow{u}_{\bot}$ be an arbitrary column of computed basis $U_{\bot}$. Select $\overrightarrow{w}_{k,N,p_k}^{(n_k-1)}$ such that $p_k>0$. We have
\begin{equation}
\overrightarrow{u}_{\bot}^H\overrightarrow{w}_{k,N,p_k}^{(n_k-1)} = [\overrightarrow{u}_{\bot}^H\overrightarrow{w}_{k,N}^{(n_k-1)}, \overrightarrow{u}_{\bot}^H\overrightarrow{w}_{k,N}^{(n_k)}] = [0,0]
\end{equation}
\noindent Equivalently, 
\begin{equation}
[\overrightarrow{u}_{\bot}^H\overrightarrow{w}_{k,N}^{(n_k-1)}, \overrightarrow{u}_{\bot}^H(r_k\overrightarrow{w}_{k,N}^{(n_k)})] = [0,0]
\end{equation}
\noindent which implies
\begin{equation}
\label{break}
\overrightarrow{u}_{\bot}[n_k-1] = \overrightarrow{u}_{\bot}^H\overrightarrow{w}_{k,N}^{(n_k-1)} - \overrightarrow{u}_{\bot}^H(r_k\overrightarrow{w}_{k,N}^{(n_k)}) = 0
\end{equation}
\noindent Since $\overrightarrow{u}_{\bot}$ is an arbitrary column of $U_{\bot}$, then every element in row $n_k-1$ of $U_{\bot}$ is zero. In this case,
\begin{equation}
\label{zero_row}
(\overrightarrow{w'}_N^{(n_k-1)})^H\times U_{\bot} = z\times(\overrightarrow{w'}_N^{(n_k)})^H\times U_{\bot} 
\end{equation}
\noindent Substituting~(\ref{zero_row}) in~(\ref{eqSet}) we get
\begin{equation}
\label{equivEq}
J(z)^{n_k-1} = z\times J(z)^{n_k}\times z^H = 0
\end{equation}
\noindent Thus equations $J(z)^{n_k-1} = 0$ and $J(z)^{n_k-1} = 0$ are equivalent, and every solution of one of these equations is also a solution of the other. This is true whether this solution is a complex exponential on the unit circle corresponding to an aligned or misaligned transmission, or a complex root  of the high-order polynomial~(\ref{equivEq}) not on the unit circle. Therefore, the receiver can easily identify when a misaligned transmission occurs by simply searching for zero rows in $U_{\bot}$. However, if for example an aligned transmission starts in the same time slot as the misaligned transmission, the receiver cannot identify which of these two specific transmissions is the aligned one and which is the misaligned. This is because both unit-magnitude roots $r_k$ will be duplicated.\\

\noindent We now modify the transmission algorithm. We still assume that any packet arrives at the start of a symbol duration. However, regardless whether a transmitter is aligned or misaligned, every transmitter $k$ sends its first packet as $1\times r_k^0\overrightarrow{s}_k$. If a retransmission is necessary, transmitter $k$ sends $2\times r_k^1 \overrightarrow{s}_k$. In its third time, the transmitter sends $1\times r_k^2 \overrightarrow{s}_k$, and so on. This is to say that each transmitter sends its packets weighted as before except for an extra factor of $2$ every other transmission (at even indices). Define $\overrightarrow{w}_{k,N,2}$ such that
\begin{equation}
\overrightarrow{w}_{k,N,2}[n] = \begin{cases}
\overrightarrow{w}_{k,N}[n],\quad\quad n \text{   odd}\\
2\times\overrightarrow{w}_{k,N}[n],\quad\quad n \text{   even} 
\end{cases}
\end{equation}
\noindent Let
\begin{equation}
\overrightarrow{w}_{k,N,p_k,2}^{(n_k-1)} = \begin{cases}
\overrightarrow{w}_{k,N,2}^{(n_k-1)},\quad\quad\quad p_k = 0\\
\big[\overrightarrow{w}_{k,N,2}^{(n_k-1)},\overrightarrow{w}_{k,N,2}^{(n_k)}\big],~~0<p_k<P
\end{cases}
\end{equation}
\noindent The collected matrix $Y_N$ becomes
\begin{equation}
\label{eqY_Ndetailed3}
\begin{split}
Y_N &= \begin{pmatrix}
\overrightarrow{w}_{1,N,p_1,2}^{(n_1-1)} & \hdots & \overrightarrow{w}_{K,N,p_K,2}^{(n_K-1)}
\end{pmatrix} \times S_p + \mathcal{N}_{N,P}\\
&= W_{N,p,2}^{(n-1)}\times S_p + \mathcal{N}_{N,P}
\end{split}
\end{equation}
\noindent Define
\begin{equation}
\overrightarrow{w'}_{N,2}[n] = \begin{cases}
\overrightarrow{w'}_{N}[n],\quad\quad n \text{   odd}\\
2\times\overrightarrow{w'}_{N}[n],\quad\quad n \text{   even} 
\end{cases}
\end{equation}
\noindent The receiver solves the modified set of independent equations
\begin{equation}
\label{eqSet2}
J(z)_2^{n_k-1} = (\overrightarrow{w'}_{N,2}^{(n_k-1)})^H\times U_{\bot} U_{\bot}^H\times \overrightarrow{w'}_{N,2}^{(n_k-1)} = 0
\end{equation} 
\noindent for $1\leq n_k\leq N-1$. Relations~(\ref{break}),~(\ref{zero_row}) and~(\ref{equivEq}) are no more satisfied. $U_{\bot}$ no more has zero rows. The only solutions of~(\ref{eqSet2}) with unity-magnitude that will show as duplicates are those corresponding to misaligned transmissions. The receiver is thus able to construct $W_{N,p,2}^{(n-1)}$ and recover $S_p$ as
\begin{equation}
\hat{S}_p = ((W_{N,p,2}^{(n-1)})^H W_{N,p,2}^{(n-1)})^{-1} (W_{N,p,2}^{(n-1)})^H Y_N
\end{equation}
\noindent Generally, $0\leq p_k < P$ may be a non-integer. In this case $W_{N,p,2}^{(n-1)}$ in~(\ref{eqY_Ndetailed3}) is unaltered, and thus neither the receiver nor the transmitters modify their algorithms. However, packet $\overrightarrow{s}_k$ of $P$ symbols in $S_p$ now extends over $P+1$ symbol durations. Therefore $S_p$ will hold $P+1$ virtual symbols in place of the original $P$ symbols. The receiver is able to recover these $P+1$ virtual symbols using the algorithm illustrated above. In addition, the receiver reconstructs the analog signal from the $P+1$ virtual symbols. In principle only $P\tau$ seconds of the analog signal hold symbol information, and the receiver resamples for the $P$ actual symbols. Note that in the analog domain, the signal reconstructed by the receiver is simply a phase-shifted version of the analog signal first issued by transmitter $k$. Yet, within a symbol duration a virtual symbol is an intermixture of two actual consecutive symbols (except at the edges) and applying a phase shift directly in the symbol domain generally does not work. Note also that unless $p_k$ is an integer or can be approximated by an integer, packet $\overrightarrow{s}_k$ always occupies two rows of $S_p$ equivalent to two time slots. This is because $\overrightarrow{s}_k$ has $P+1$ virtual symbols while $S_p$ is only $P$ columns wide.\\

\noindent In the worst case when no single transmission is aligned, every packet in $S_p$ overlaps with two time slots and the asymptotic throughput is
\begin{equation}
\lim_{K\rightarrow \infty}\frac{K}{N} \geq 50\%
\end{equation}
\noindent Again, the algorithm in Section~\ref{sec4} is a special case of the one in this section and is superseded as it only treats the scenario where all transmitters are aligned.

\section{Static Channel Effects}~\label{sec6}
Let $q_k$ be a complex number that represents static channel effects such as pathloss on the link between transmitter $k$ and the receiver. Define the diagonal matrix $D = \text{diag}(q_1,\dots,q_K)$. If transmitter $k$ is misaligned then $q_k$ shows twice in $D$. $Y_N$ can be expressed as
\begin{equation}
Y_N = W_{N,p,2}^{(n-1)}\times D\times S_p + \mathcal{N}_{N,P}
\end{equation}
\noindent The receiver applies the algorithm of Section~\ref{sec5} and recovers $D\times S_p$. Using the preamble bits of each packet $\overrightarrow{s}_k$, the receiver computes $D$ and extracts $S_p$.

\section{Fading Channel Model}~\label{sec7}
Assume a general channel model as described in~Section~\ref{sec2}. $Y_N$ becomes
\begin{equation}
Y_N = W_{N,p,2,h}^{(n-1)}\times D\times S_p + \mathcal{N}_{N,P}
\end{equation}
\noindent where $W_{N,p,2,h}^{(n-1)}$ is the same as $W_{N,p,2}^{(n-1)}$ except that each element in the $n^{\text{th}}$ row and the $k^{th}$ vertical block of $W_{N,p,2}^{(n-1)}$ is scaled by $h_k(n)$. The receiver applies the same algorithm of Sections~\ref{sec6} and~\ref{sec7} but searches for $K$ unit-magnitude complex solutions within $K$ sets of equations. Each set is basically the same as~(\ref{eqSet2}) except that in the $k^{\text{th}}$ set of equations, $\overrightarrow{w'}_{N,2}^{(n_k-1)}$ is replaced by steering vector $\overrightarrow{w'}_{N,2,h}^{(n_k-1)}$ where $\overrightarrow{w'}_{N,2,h}^{(n_k-1)}[n] = h_k(n)\times  \overrightarrow{w'}_{N,2}^{(n_k-1)}[n]$, $1\leq n\leq N$. Note that there should be a correspondence between the index of the set of equations $k$ and the found roots $r_k$, i.e. $r_{k'}$ cannot be detected as one of the $K$ solutions upon solving the $k^{th}$ set of equations where $k\neq k'$. 

\section{Numerical Simulations}~\label{sec8}
In this section we show through simulations the effect of the SNR on the ability of the receiver to correctly detect the identity of the transmitters on one hand, and to correctly decode the packets of the identified transmitters on the other hand. We assume the network has 32 transmitters of which only $K=8$ are active. The 32 transmitters are assigned equally-spaced angles between $0$ and $\pi$, and the $K$ transmitters are randomly selected. The first packet from each of the $K$ transmitters arrives at $t=0$ as in Section~\ref{sec3}. Each packet is of length $P=24$, and so $S$ is a matrix of $8\times 24$ random 8-bit integers between 0 and 255. Having each symbol hold 8 bits increases the sensitivity to the SNR. All elements of S are real. We vary $\sigma^2$ on a log-scale between 1e-6 and 1e3, and we define the SNR as $\text{SNR} = 10\log_{10}(1/\sigma^2)$. We assume the noise power is equally distributed on the real and imaginary values of the received samples of $Y_N$. For each value of $\sigma^2$ we run the simulation 1000 times and compute the mean statistics.\\

\begin{figure}[ht]
\begin{center} 
\includegraphics[width=3.5 in]{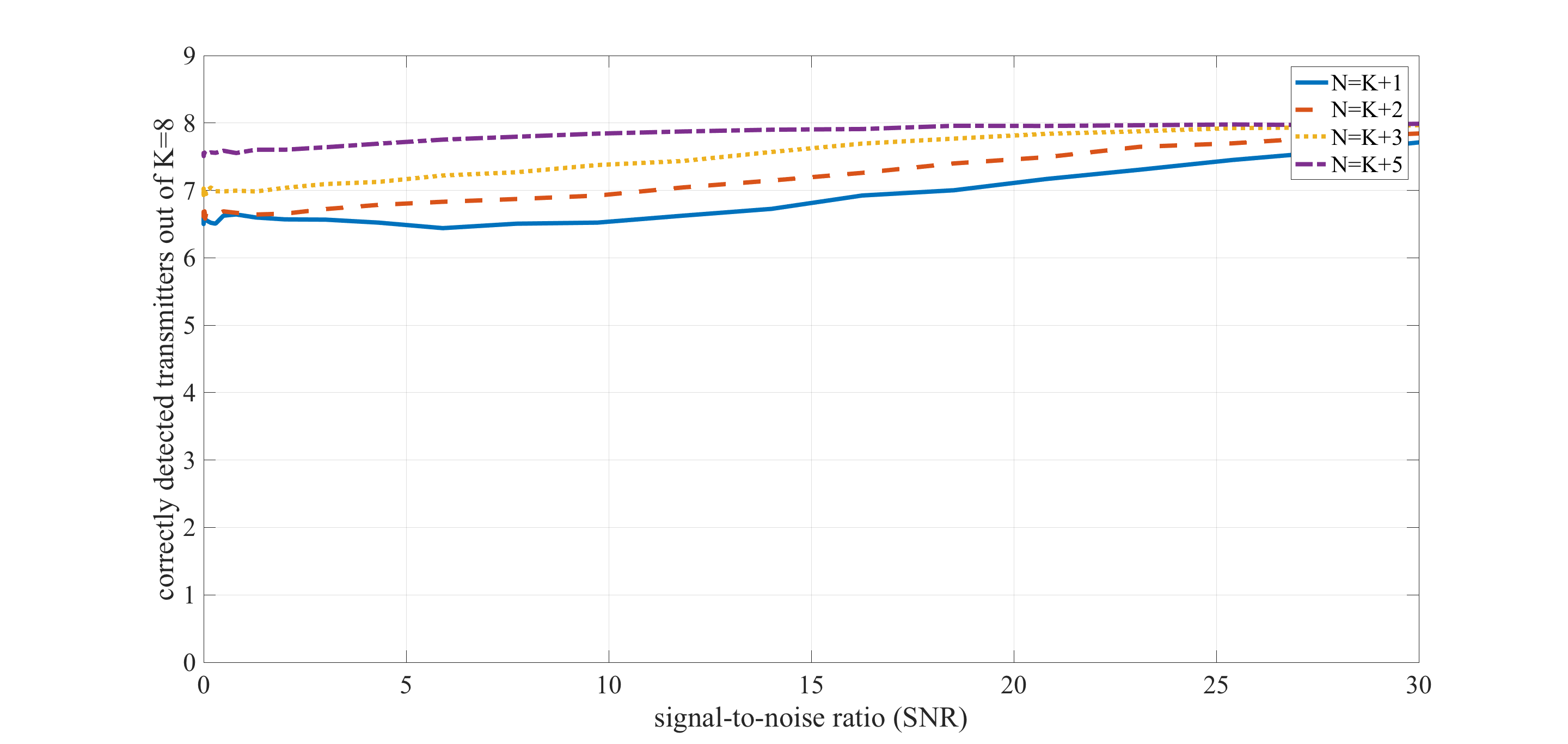}
\caption{Variation of the number of correctly detected transmitters out of $K$ versus $SNR = 10\log_{10}(1/\sigma^2)$ for $N-K=1,2,3 \text{ and }5$ and aligned transmissions.}
\label{fig1}
\end{center}
\end{figure} 

\noindent In figure~\ref{fig1} we check the number of correctly detected transmitters out of $K$ within the superset of $32$ transmitters. Here we assume that the receiver knows at every SNR what threshold to use in order to decide whether a singular value of the left nullspace of $Y_N$ corresponds to noise and should be nullified. Thus, the receiver knows $K$ and the only error that might occur is that it misidentifies which $K$-selection of the $32$ transmitters is the active set. We examine the performance of the receiver when it collects $N - K = 1, 2, 3, \text{ and }5$ extra packets after detecting a full rank matrix $Y_{n = K}$. We can see that for all stopping times $N$ the number of correctly identified transmitters is increasing with the SNR. Moreover, the more vectors $\overrightarrow{y}_n$ collected for $n>K$, the more accurate the labeling of the transmitters becomes at all SNR. This is because the receiver acquires more dimensions of the noise subspace $U_{\bot}$ orthogonal to the signal-and-noise subspace $U_{\parallel}$. All $K$ transmitters are detected at high enough SNR and $N-K = 5$.

\begin{figure}[ht]
\begin{center}
\includegraphics[width=3.5 in]{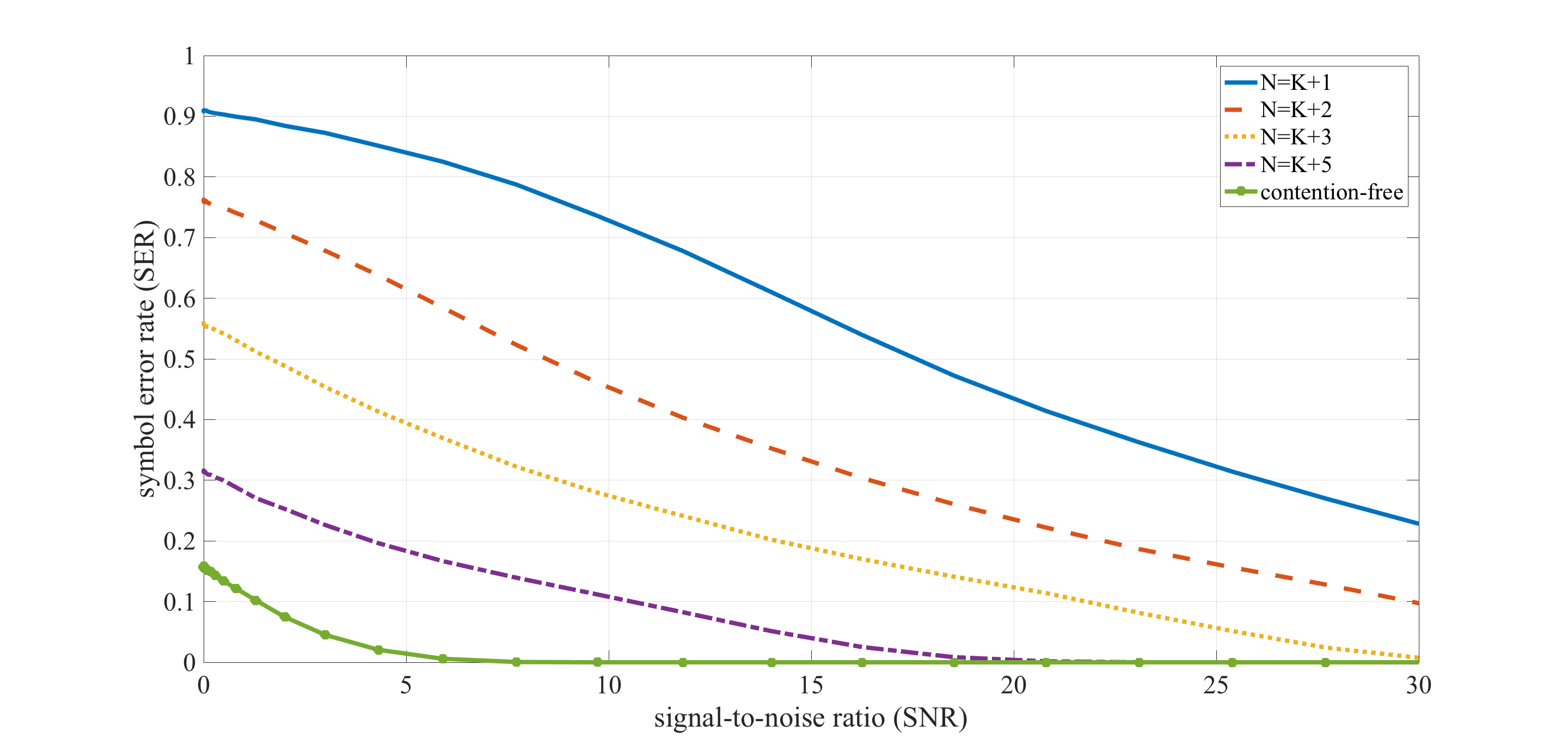}
\caption{Variation of symbol error rate versus $SNR = 10\log_{10}(1/\sigma^2)$ for $N-K=1,2,3 \text{ and }5$ and aligned transmissions.}
\label{fig2}
\end{center}
\end{figure}

\noindent We now check the symbol error rate SER of the decoded packets for the case where all $K$ transmitters are known to the receiver. The SER is always greater than or equal the bit error rate BER. As expected, this number drops for higher SNR. It also drops when $N-K$ increases. For $N-K=5$ and high SNR all symbols are decoded correctly.

\section{Conclusion}~\label{sec9}
In this paper we show how a receiver correctly extracts a desired packet when it arrives simultaneously with other desired colliding packets and undesired interference in many-to-one and many-to-many communication scenarios. The algorithm achieves full asymptotic throughput at high SNR, local CSI and slot-synchronization between each receiver and its candidate set of transmitters. Since synchronization is commonly difficult to achieve in practice, the algorithm also resolves unsynchronized transmissions at a lower throughput that is still higher than collision resolution protocols and interference alignment techniques. This should give insight into the design of networks as it is best if only nodes commonly heard at a receiver are synchronized with that receiver. Moreover, it is most common that interference or collisions occur between only few nodes. Thus synchronization among few nodes becomes feasible and the algorithm achieves high performance. In the simulations we show that high SNR is important. This should not be problematic since interfering high-power signals can be resolved. At a given SNR, performance of the receiver is always improved by collecting extra few packets.    

\small

\end{document}